\documentclass[twocolumn,prb]{revtex4}
\usepackage{amsfonts}
\usepackage[T1]{fontenc}
\usepackage{amsmath,amsbsy,amssymb,graphicx}
\usepackage{times}

\let\mathbf=\boldsymbol
\input{tcilatex}

\begin{document}

\title{ Electric-circuit simulation of the Schr\"{o}dinger equation\\
and non-Hermitian quantum walks }
\author{Motohiko Ezawa}
\affiliation{Department of Applied Physics, University of Tokyo, Hongo 7-3-1, 113-8656,
Japan}

\begin{abstract}
Recent progress has witnessed that various topological physics can be
simulated by electric circuits under alternating current. However, it is
still a nontrivial problem if it is possible to simulate the dynamics
subject to the Schr\"{o}dinger equation based on electric circuits. In this
work, we reformulate the Kirchhoff law in one dimension in the form of the
Schr\"{o}dinger equation. As a typical example, we investigate quantum walks
in $LC$ circuits. We also investigate how quantum walks are different in
topological and trivial phases by simulating the Su-Schrieffer-Heeger model
in electric circuits. We then generalize them to include dissipation and
nonreciprocity by introducing resistors, which produce non-Hermitian
effects. We point out that the time evolution of one-dimensional quantum
walks is exactly solvable with the use of the generating function made of
the Bessel functions.
\end{abstract}

\maketitle

Electric circuits have demonstrated their usefulness in the field of
condensed-matter physics since they can simulate various topological physics%
\cite%
{ComPhys,TECNature,Garcia,Hel,Lu,EzawaTEC,Hofmann,Research,EzawaLCR,EzawaSkin,EzawaChern,EzawaMajo,HelSkin,Zeng,Jiang,Lee}%
. It has been proved that the circuit Laplacian and the tight-binding
Hamiltonian have a one-to-one correspondence when an alternating current is
applied\cite{TECNature,ComPhys}. It is yet an open problem whether the
dynamics governed by the Schr\"{o}dinger equation can be simulated by
electric circuits. A simplest dynamical problem would be a one-dimensional
quantum walk, which we wish to explore.

Quantum walk is a diffusion process governed by the Schr\"{o}dinger equation%
\cite{Aha,Farhi,Amba,Andra,Kempe,Rud}. As a function of time, its standard
deviation spreads linearly and faster than a classical random walk which
spreads proportional to the square root of time\cite{ben,Konno}. Quantum
walk is a basic concept in quantum information processes including quantum
search\cite{Sze,ChildA}, universal quantum computation\cite{Child,
ChildScience} and quantum measurement\cite{Bian}. It is realized in photonic
lattice\cite{Peru,Guzik,White,Xiao}, wave guide\cite{Hagai} and
nuclear-magnetic-resonance\cite{Du,Rud2}.

In this paper, first we demonstrate the mathematical equivalence between the
telegrapher equation and the Schr\"{o}dinger equation. It implies that any
solution of the telegrapher equation is given by the wave function of the
Schr\"{o}dinger equation, although they may describe different physical
objects. Conversely, the dynamics governed by the Schr\"{o}dinger equation
can be simulated by electric circuits. Second, as an explicit example, we
solve the telegrapher equation analytically to simulate a quantum walker in
electric circuits. Third, we investigate how quantum walks are different in
topological and trivial phases. For this purpose we propose an
electric-circuit simulation of the Su-Schrieffer-Heeger (SSH) model.
Topological and trivial phases are well distinguished by the time evolution
of a quantum walk starting from the edge. Finally, we study a nonreciprocal
non-Hermitian quantum walk, where it is found that a quantum walker linearly
displaces while the variance increases only linearly as a function of time.
It is highly contrasted to a reciprocal quantum walk.

\begin{figure}[t]
\centerline{\includegraphics[width=0.48\textwidth]{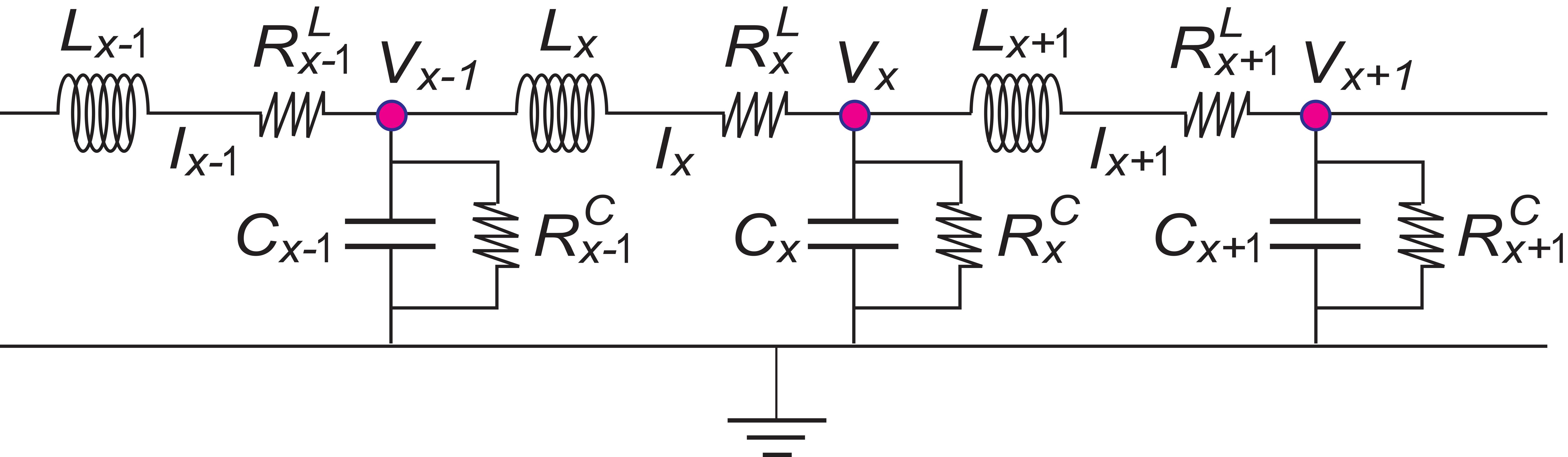}}
\caption{Illustration of an electric circuit realizing an inhomogeneous
telegrapher equation.}
\label{FigIllust}
\end{figure}

\textit{Quantum walk based on $LC$ electric circuits:} Our system is a chain
of electric circuit shown in Fig.\ref{FigIllust}. It describes the
telegrapher equation provided the system is homogeneous. An inhomogeneous
circuit is constructed by choosing the sample parameters different depending
on the position in a chain. An electric circuit is characterized by the
Kirchhoff laws (Fig.\ref{FigIllust}),%
\begin{align}
L_{x}\frac{d}{dt}I_{x} =&V_{x-1}-V_{x}-R_{x}^{L}I_{x}.  \label{EqB} \\
C_{x}\frac{d}{dt}V_{x} =&I_{x}-I_{x+1}-V_{x}/R_{x}^{C},  \label{EqA}
\end{align}%
The first equation is the Kirchhoff voltage law with respect to the voltage
difference between two nodes $V_{x}$ and $V_{x-1}$, which is equal to the
voltage drop by the resistor $R_x$ and the inductive electromotive force by
the inductor $L_x$. The second equation is the Kirchhoff current law with
respect to the current conservation at one node $V_{x}$, where the current
flows to the ground via the the conductor $C_x$ and the resister $R_x^C$ in
paralell. They are combined into a second-order differential equation by
deleting $I$ or $V$ in the standard treatment\cite{Rosen}.

We first analyze the homogeneous case such that $C_{x}=C$, $L_{x}=L$, $%
R_{x}^{L}=R^{L}$ and $R_{x}^{C}=R^{C}$. We make a scale transformation%
\begin{equation}
\mathcal{V}_{k}=V_{k},\qquad \mathcal{I}_{k}=\sqrt{\frac{L}{C}}I_{k},
\label{EqC}
\end{equation}%
so that $\mathcal{V}_{k}$ and $\mathcal{I}_{k}$ have the same dimension,
where $\sqrt{L/C}I_{k}$ is the voltage drop per unit length. The set of
equations (\ref{EqB}) and (\ref{EqA}) are reformulated in the form of the
Schr\"{o}dinger equation, 
\begin{equation}
i\partial _{t}\psi _{k}=\mathcal{H}\left( k\right) \psi _{k},
\label{SchroeEq}
\end{equation}%
with the wave function $\psi _{k}=\left( \mathcal{I}_{k},\mathcal{V}%
_{k}\right) ^{t}$, and the Hamiltonian 
\begin{equation}
\mathcal{H}\left( k\right) =\left( 
\begin{array}{cc}
-i\frac{R^{L}}{L} & -\frac{i}{\sqrt{LC}}\left( 1-e^{-ik}\right) \\ 
\frac{i}{\sqrt{LC}}\left( 1-e^{ik}\right) & -i\frac{1}{CR^{C}}%
\end{array}%
\right) .
\end{equation}%
This Hamiltonian is non-Hermitian due to the diagonal resister terms $%
-iR^{C} $ and $-iR^{L}$. The "energy spectrum" is given by%
\begin{equation}
E=-i\frac{R^{L}/L+1/CR^{C}}{2}\pm \sqrt{\frac{4}{LC}\sin ^{2}k-\left( \frac{%
R^{L}}{L}-\frac{1}{CR^{C}}\right) ^{2}}.
\end{equation}%
The dynamics is solved as $\psi _{k}\left( t\right) =e^{i\mathcal{H}\left(
k\right) t}\psi _{k}\left( 0\right) $.

\begin{figure}[t]
\centerline{\includegraphics[width=0.48\textwidth]{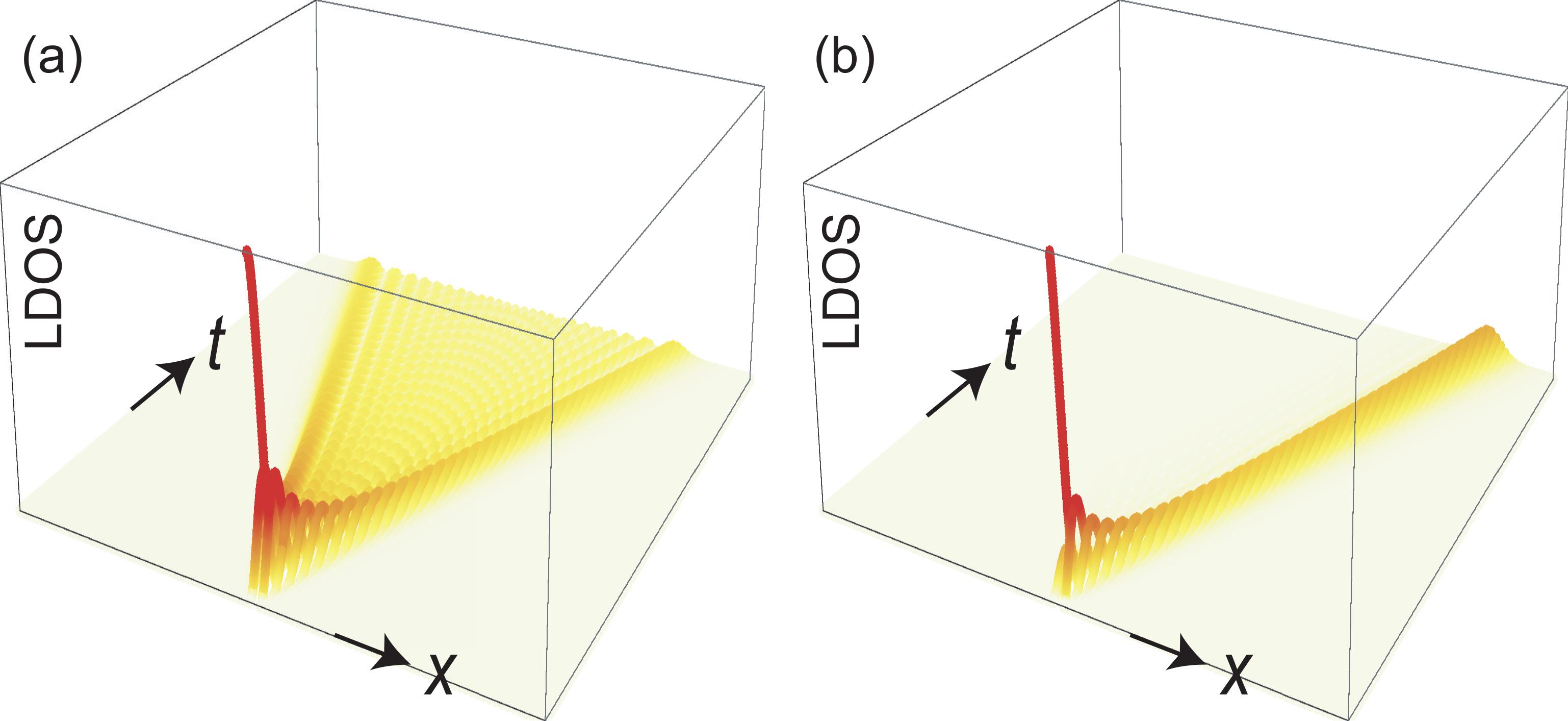}}
\caption{(a) Time evolution of a quantum walk in electric circuits. (b) Time
evolution of a quantum walk with nonreciprocity ($\protect\gamma =1.25$) and
dissipation ($R=0.4$). The vertical axis is the LDOS constructed from (%
\protect\ref{Voltage}), which is the square of voltage. We have set $%
L_{x}=C_{x}=1$.}
\label{FigWalk}
\end{figure}

For simplicity we set $R^{L}/L=1/CR^{C}=R$. The telegrapher equation is a
set of equations (\ref{EqA}) and (\ref{EqB}), which is converted to 
\begin{equation}
i\frac{d}{dt}\psi _{x}=\frac{i}{\sqrt{LC}}\psi _{x-1}-iR\psi _{x}-\frac{i}{%
\sqrt{LC}}\psi _{x+1},  \label{SchEqA}
\end{equation}%
by transforming (\ref{SchroeEq}) into the real space, where we have defined%
\begin{equation}
\psi _{x}=\left( \cdots ,\mathcal{I}_{x-1},\mathcal{V}_{x-1},\mathcal{I}_{x},%
\mathcal{V}_{x},\mathcal{I}_{x+1},\mathcal{V}_{x+1},\cdots \right) ^{t}.
\label{Voltage}
\end{equation}%
Let us show that transient phenomena described by (\ref{SchEqA}) together
with appropriate initial conditions are mathematically equivalent to the
dynamics of quantum walkers.

We start with a quantum walker starting from $x=0$ at $t=0$. Namely, we
solve (\ref{SchEqA}) by imposing the initial condition $\psi _{x}=\delta
_{x0}$ at $t=0$. The analytic solution is obtained as%
\begin{equation}
\psi _{x}=e^{-Rt}J_{\left\vert x\right\vert }\left( \frac{2}{\sqrt{LC}}%
t\right) ,  \label{QWakerSol}
\end{equation}%
where $J_{x}$ is the Bessel function. We show the time evolution of the
eigenstate in Fig.\ref{FigWalk}(a). The eigenstate is observable by
measuring the voltage and the current.

We discuss analytically how the wave packet describing the quantum walker
spreads throughout the lattice as shown in Fig.\ref{FigWalk}(a). We define a
generating function by\cite{Konno}%
\begin{equation}
G\left( k\right) =\sum_{x=-\infty }^{\infty }\left\vert \psi _{x}\left(
t\right) \right\vert ^{2}e^{kx}.
\end{equation}%
Using the sum formula of the Bessel function,%
\begin{equation}
\sum_{x=-\infty }^{\infty }J_{\left\vert x\right\vert }^{2}\left( t\right)
e^{kx}=I_{0}\left( t\sqrt{2\left( \cosh k-1\right) }\right) ,
\end{equation}%
we find%
\begin{equation}
G\left( k\right) =e^{-2Rt}I_{0}\left( 2t\sqrt{\frac{2\left( \cosh k-1\right) 
}{LC}}\right) ,
\end{equation}%
where $I_{0}$ is the modified Bessel function. The $n$-th moment is
calculated as%
\begin{equation}
\left\langle x^{n}\right\rangle =\lim_{k\rightarrow 0}\frac{d^{n}G\left(
k\right) }{dk^{n}}.
\end{equation}%
The total density decreases as 
\begin{equation}
\left\langle 1\right\rangle =G\left( 0\right) =e^{-2Rt}
\end{equation}%
in the presence of the dissipation $R$. Indeed, we obtain $\sum_{x=-\infty
}^{\infty }\left\vert \psi _{x}\left( t\right) \right\vert ^{2}=1$ for $R=0$%
. The mean position is $\left\langle x\right\rangle =0$, while the variance
is 
\begin{equation}
\left\langle x^{2}\right\rangle =\frac{4t^{2}}{LC}e^{-2Rt}.
\end{equation}%
Hence, in the absence of the dissipation ($R=0$), the variance diffuses
quadratically or the standard deviation increases linearly as a function of
time. This is a manifestation of a quantum walk\cite{ben,Konno}.

\begin{figure}[t]
\centerline{\includegraphics[width=0.48\textwidth]{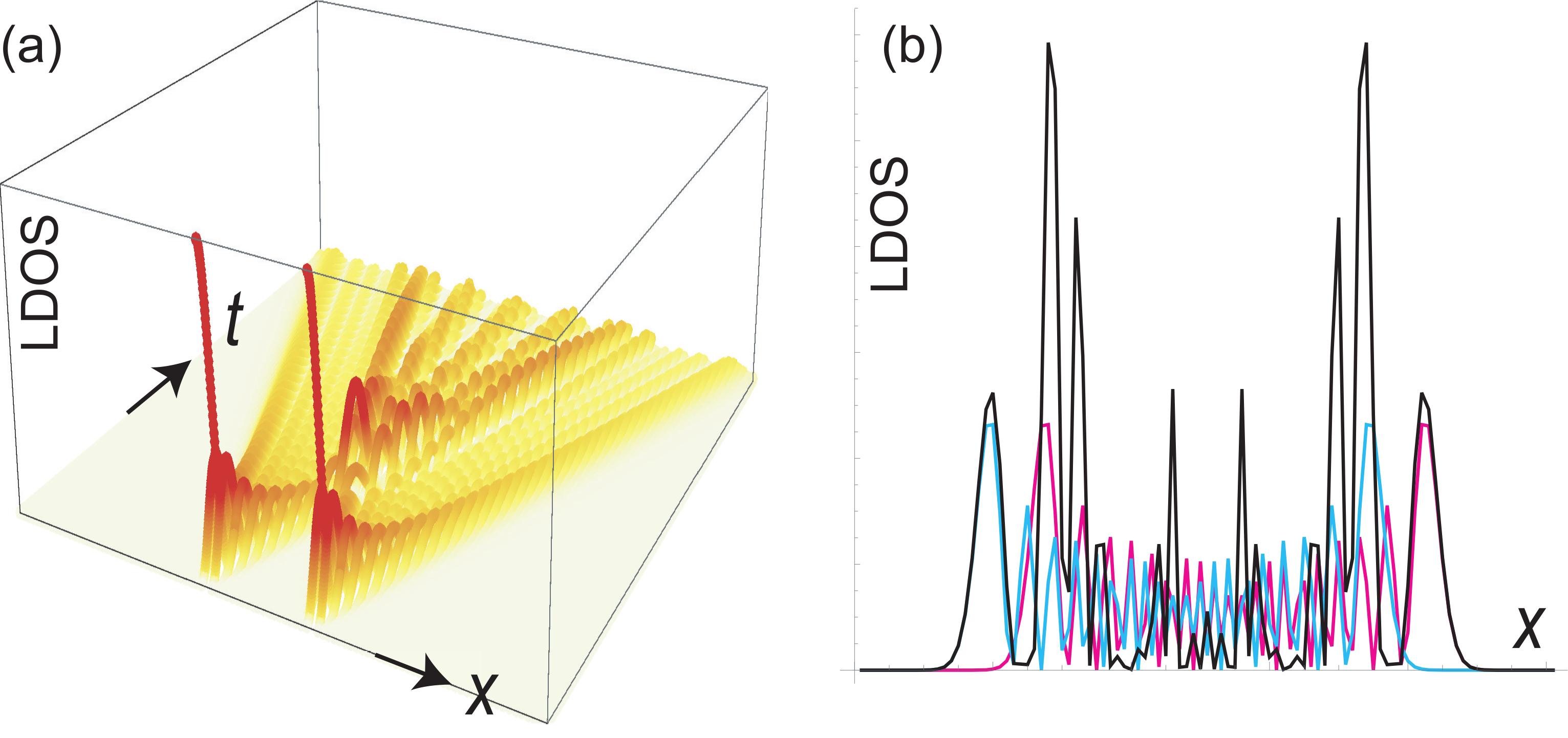}}
\caption{Electric-circuit simulation of interference experiment. (a) Time
evolution of two quantum walkers starting from two points. (b) Absolute
value of the LDOS at a fixed time $t$. The magenta (cyan) curve corresponds
to the probability to find the right (left) walker at a certain point, while
the black curve corresponds to the probability to find a walker at a certain
point. The black curve clearly forms an interference pattern. We have set $%
L=C=1$, $R=0$ and $t=30$.}
\label{FigYoung}
\end{figure}

\textit{Interference experiment:} We analyze the problem of two quantum
walkers. Let their starting points be $x\pm x_{0}$ at $t=0$. The eigen
function is simply given by a linear superposition of two eigenstates of the
type (\ref{QWakerSol}),%
\begin{equation}
\psi _{x}=\psi _{x}^{+}+\psi _{x}^{-},
\end{equation}%
where%
\begin{equation}
\psi _{x}^{\pm }=e^{-Rt}J_{\left\vert x\pm x_{0}\right\vert }\left( \frac{2}{%
\sqrt{LC}}t\right) .
\end{equation}%
We show the absolute value of $\psi _{x}$ for a fixed time in Fig.\ref%
{FigYoung}(b). An interference pattern is clearly observed.

\textit{Quantum walk in inhomogeneous system:} We generalize the results to
an inhomogeneous system. By making a spatial dependent scale transformation 
\begin{equation}
V_{x}=\alpha _{x}\mathcal{V}_{x},\qquad I_{x}=\beta _{x}\mathcal{I}_{x}
\label{ScaleTr}
\end{equation}%
in (\ref{EqB}) and (\ref{EqA}), we obtain equations%
\begin{align}
i\frac{d}{dt}\mathcal{I}_{x}=& \frac{i\alpha _{x-1}}{\beta _{x}L_{x}}%
\mathcal{V}_{x-1}-\frac{iR_{x}^{L}}{L_{x}}\mathcal{I}_{x}-\frac{i\alpha _{x}%
}{\beta _{x}L_{x}}\mathcal{V}_{x}. \\
i\frac{d}{dt}\mathcal{V}_{x}=& \frac{i\beta _{x}}{\alpha _{x}C_{x}}\mathcal{I%
}_{x}-\frac{i}{C_{x}R_{x}^{C}}\mathcal{V}_{x}-\frac{i\beta _{x+1}}{\alpha
_{x}C_{x}}\mathcal{I}_{x+1},
\end{align}%
These equations lead to a non-Hermitian Hamiltonian.

By choosing $\alpha _{x}=\sqrt{C_{1}/C_{x}}$ and $\beta _{x}=\sqrt{%
C_{1}/L_{x}}$ in (\ref{ScaleTr}), the set of equations become 
\begin{align}
i\frac{d}{dt}\mathcal{I}_{x}=& i\sqrt{\frac{1}{L_{x}C_{x-1}}}\mathcal{V}%
_{x-1}-\frac{iR_{x}^{L}}{L_{x}}\mathcal{I}_{x}-i\sqrt{\frac{1}{L_{x}C_{x}}}%
\mathcal{V}_{x}. \\
i\frac{d}{dt}\mathcal{V}_{x}=& i\sqrt{\frac{1}{L_{x}C_{x}}}\mathcal{I}_{x}-%
\frac{i}{C_{x}R_{x}^{C}}\mathcal{V}_{x}-i\sqrt{\frac{1}{L_{x+1}C_{x}}}%
\mathcal{I}_{x+1},
\end{align}%
When we set $R_{x}^{L}/L_{x}=1/C_{x}R_{x}^{C}=R$ for simplicity, the
corresponding tight-binding Hamiltonian has a particularly simple form, 
\begin{equation}
H=\sum_{x}t_{x}\left( \left\vert \psi _{x}\right\rangle \left\langle \psi
_{x+1}\right\vert +\left\vert \psi _{x+1}\right\rangle \left\langle \psi
_{x}\right\vert \right) -iR\left\vert \psi _{x}\right\rangle \left\langle
\psi _{x}\right\vert ,  \label{HamilTB}
\end{equation}%
with $t_{2x-1}=1/\sqrt{L_{x}C_{x}}$ and $t_{2x}=1/\sqrt{L_{x+1}C_{x}}$.
Here, $t_{x}$ represents the hopping parameter between two sites $x$ and $x+1
$. The inverse solutions are given by%
\begin{equation}
\frac{L_{x}}{L_{1}}=\left( \frac{\prod\limits_{j=1}^{x-1}t_{2j-1}}{%
\prod\limits_{j=1}^{x-1}t_{2j}}\right) ^{2},\qquad \frac{C_{x}}{C_{1}}%
=\left( \frac{\prod\limits_{j=1}^{x-1}t_{2j+1}}{\prod%
\limits_{j=1}^{x-1}t_{2j}}\right) ^{2}.  \label{LCn}
\end{equation}%
Consequently, it is possible to arrange capacitors $C_{x}$ and inductors $%
L_{x}$ to reproduce various tight-binding models with arbitrary hopping
parameters $t_{x}$.

\begin{figure}[t]
\centerline{\includegraphics[width=0.48\textwidth]{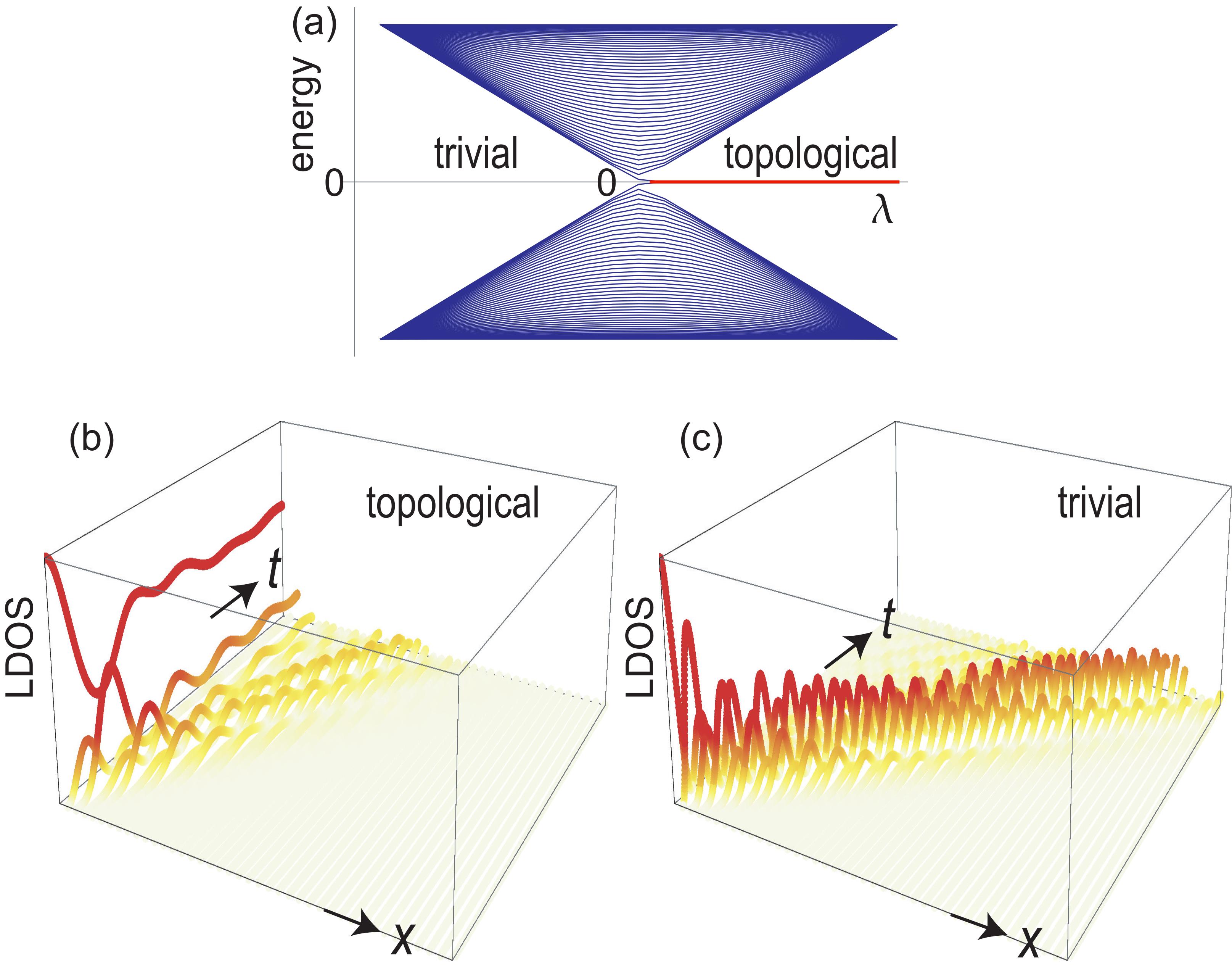}}
\caption{(a) "Energy spectrum" as a function of $\protect\lambda $. There
are "zero-energy" edge states indicated by a red line in the topological
phase ($\protect\lambda >0$), while there are no edge states in the trivial
phase ($\protect\lambda <0$). (b) Time evolution of a quantum walk in the
topological phase ($\protect\lambda =0.5$), and (c) that in the trivial
phase ($\protect\lambda =-0.5$). A quantum walk does not diffuse for the
topological phase, while it diffuses for the trivial phase.}
\label{FigCWalk}
\end{figure}

\textit{Quantum walks in topological and trivial phases:} We proceed to
investigate how quantum walks are different in topological and trivial
phases. The simplest model possessing these phases is given by the SSH model%
\cite{Simon}. The SSH model is given by the Hamiltonian (\ref{HamilTB})
together with $R=0$ and%
\begin{equation}
t_{x}=t+\left( -1\right) ^{x}\lambda .
\end{equation}%
The electric circuit is constructed by choosing inductors and capacitors
satisfying (\ref{LCn}). The energy spectrum is shown as a function of $%
\lambda $ in Fig.\ref{FigCWalk}(a). There are "zero-energy" edge states for $%
\lambda >0$, signaling that the system is topological, while the system is
trivial for $\lambda <0$ with no edge states. The edges are said topological
for $\lambda >0$. These two phases are clearly distinguishable by examining
quantum walks. Let a quantum walker start from one of the edges. Namely, we
consider the initial state chosen to be perfectly localized at one edge. In
the topological phase, a nonzero local density of state (LDOS) remains at
the edge as shown in Fig.\ref{FigCWalk}(b), while the LDOS rapidly decreases
for the trivial phase as shown in Fig.\ref{FigCWalk}(c).

These behaviors are understood analytically as follows. By expanding the
initial state in terms of the eigenstates as%
\begin{equation}
\psi _{x}^{\text{ini}}=\sum_{j}c_{j}\psi _{x}^{\left( j\right) },
\end{equation}%
the dynamics is given by%
\begin{equation}
\psi _{x}\left( t\right) =\sum_{j}c_{j}e^{iE_{j}}\psi _{x}^{\left( j\right)
},
\end{equation}%
where $E_{j}$ is the $j$-th energy, and $\psi _{x}^{\left( j\right) }$ is
the eigenstate with the energy $E_{j}$. For the topological phase the
coefficient $c_{j}$ has the largest value for the edge state, which has no
dynamics. It results in a nonzero LDOS at the edge as in Fig.\ref{FigCWalk}%
(b). On the other hand, there is no dominant $c_{j}$ for the trivial states,
which results in the rapid spread of the initial state in Fig.\ref{FigCWalk}%
(c). 
\begin{figure}[t]
\centerline{\includegraphics[width=0.48\textwidth]{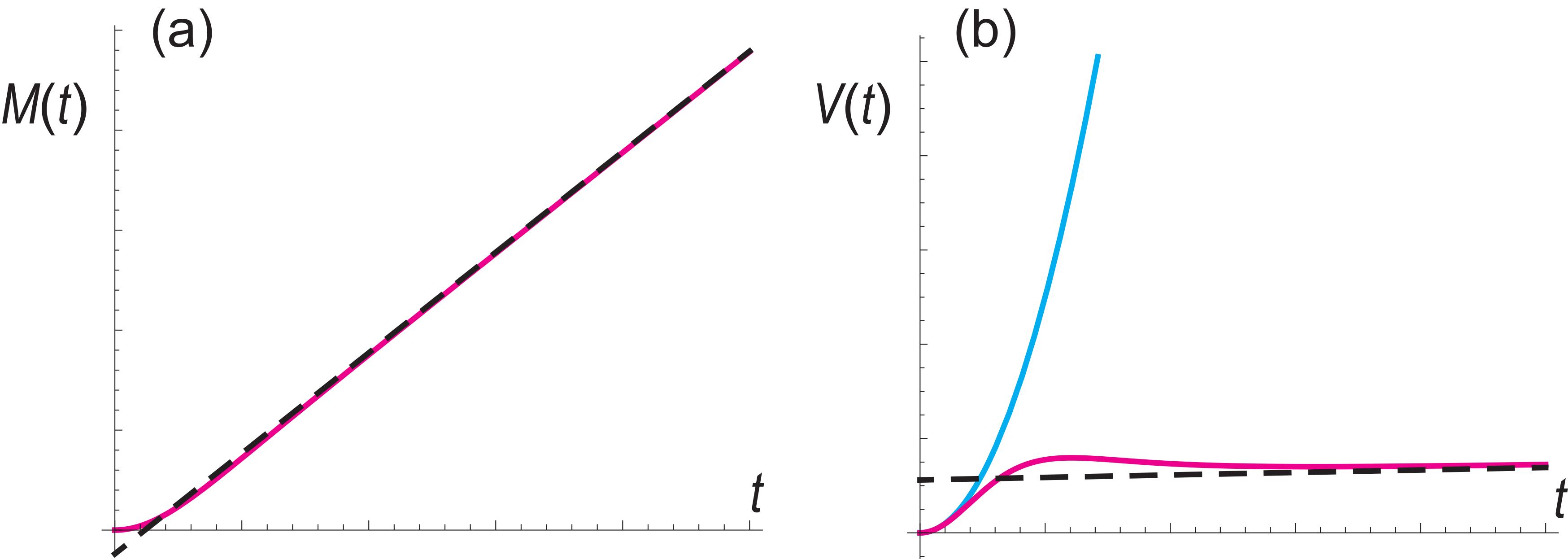}}
\caption{(a) Time evolution of the mean value $M(t)$ in electric circuits.
(b) Time evolution of the variance $V(t)$. Cyan curves represent a
reciprocal quantum walk, while magenta curves represent a nonreciprocal with 
$\protect\gamma =1.1$ quantum walk. Black dotted lines are asymptotic
formula for $t\rightarrow \infty $.}
\label{FigEV}
\end{figure}

\textit{Non-Hermitian nonreciprocal quantum walk:} Next, by choosing 
\begin{equation}
\alpha _{x}=\gamma ^{-2x}\sqrt{\frac{C_{1}}{C_{x}}},\qquad \beta _{x}=\gamma
^{-2x+1}\sqrt{\frac{C_{1}}{L_{x}}}
\end{equation}%
in (\ref{ScaleTr}), we construct a non-Hermitian nonreciprocal model\cite%
{Hatano,UedaPRX,SkinTop,EzawaSkin}, 
\begin{equation}
H=\sum_{x}t_{x}\left( \gamma \left\vert \psi _{x}\right\rangle \left\langle
\psi _{x+1}\right\vert +\frac{1}{\gamma }\left\vert \psi _{x+1}\right\rangle
\left\langle \psi _{x}\right\vert \right) -iR\left\vert \psi
_{x}\right\rangle \left\langle \psi _{x}\right\vert ,
\end{equation}%
where the parameter $\gamma $\ represents the nonreciprocity. The
telegrapher equation is given by%
\begin{equation}
i\frac{d}{dt}\psi _{x}=\frac{i\gamma }{\sqrt{LC}}\psi _{x-1}-iR\psi _{x}-%
\frac{i}{\gamma \sqrt{LC}}\psi _{x+1}.
\end{equation}%
We find an analytic solution%
\begin{equation}
\Psi _{x}\left( t\right) =\gamma ^{x}e^{-Rt}J_{\left\vert x\right\vert
}\left( \frac{2}{\sqrt{LC}}t\right) .
\end{equation}%
The generating function is%
\begin{equation}
G\left( k\right) =e^{-2Rt}I_{0}\left( 2t\sqrt{\frac{\gamma
^{2}e^{k}+e^{-k}/\gamma ^{2}-2}{LC}}\right) .
\end{equation}%
The total LDOS reads 
\begin{equation}
\sum_{x=-\infty }^{\infty }\left\vert \psi _{x}\left( t\right) \right\vert
^{2}=e^{-2Rt}I_{0}\left( 2t\sqrt{\frac{\gamma ^{2}+1/\gamma ^{2}-2}{LC}}%
\right) .  \label{M}
\end{equation}%
Indeed, it reproduces the result $\left\langle 1\right\rangle \equiv
\sum_{x=-\infty }^{\infty }\left\vert \psi _{n}\left( t\right) \right\vert
^{2}=1$ for $\gamma =1$ and $R=0$. The mean value is defined by $M\left(
\Psi \right) =\left\langle x\right\rangle /\left\langle 1\right\rangle $,
where we note $\left\langle 1\right\rangle \neq 1$ in general, whose
asymptotic behavior is given by%
\begin{align}
& \lim_{t\rightarrow \infty }M\left( \Psi \right)   \notag \\
=& \frac{\gamma ^{2}-1/\gamma ^{2}}{4\left( \gamma ^{2}+\frac{1}{\gamma ^{2}}%
-2\right) }\left( \frac{4t}{\sqrt{LC}}\sqrt{\gamma ^{2}+\frac{1}{\gamma ^{2}}%
-2}-1\right) .
\end{align}%
The variance is given by $V\left( \Psi \right) =\left\langle
x^{2}\right\rangle /\left\langle 1\right\rangle -M^{2}\left( \Psi \right) $,
which reads%
\begin{equation}
\lim_{t\rightarrow \infty }V\left( \Psi \right) =\frac{1}{2}\left( \frac{%
\gamma ^{2}}{\left( 1-\gamma ^{2}\right) ^{2}}+\sqrt{\gamma ^{2}+\frac{1}{%
\gamma ^{2}}-2}\frac{t}{\sqrt{LC}}\right) ,
\end{equation}%
where we have used the asymptotic formula of the modified Bessel function $%
\lim_{t\rightarrow \infty }I_{0}\left( t\right) =e^{t}/\sqrt{2\pi t}$. We
show the time evolution of the mean value and the variance in Fig.\ref{FigEV}%
. The asymptotic behaviors well reproduce the analytic results.

\textit{Discussions:} In this work we have demonstrated that the telegrapher
equation and the Schr\"{o}dinger equation are mathematically equivalent.
Consequently, their eigen functions are identical although they describe
different physical objects. It is important that the mathematical
equivalence justifies us to use the eigen function (\ref{Voltage}) in the
electric-circuit system to simulate the quantum dynamics governed by the Schr%
\"{o}dinger equation. As an explicit example, we have derived the
oscillatory pattern characteristic to a quantum walk, provided electric
circuits are appropriately designed.

We have also studied dissipative and nonreciprocal quantum walks by tuning
sample parameters. In a nonreciprocal quantum walk, the variance is
proportional to time which is smaller than that in a reciprocal quantum
walk, where it is proportional to the square of time. It will be a benefit
for future high-speed quantum search. Electric circuits have a merit that
they are easily equipped compared with other methods such as photonic,
wave-guide and nuclear-magnetic resonant systems. Furthermore, there is a
potentiality to construct integrated circuits of quantum walks.

The author is very much grateful to N. Nagaosa and E. Saito for helpful
discussions on the subject. This work is supported by the Grants-in-Aid for
Scientific Research from MEXT KAKENHI (Grants No. JP17K05490, No. JP15H05854
and No. JP18H03676). This work is also supported by CREST, JST (JPMJCR16F1).

\end{document}